\documentclass[conference]{IEEEtran}

\usepackage{graphicx}
\usepackage{amsmath}
\usepackage{amssymb}
\usepackage{booktabs}
\usepackage{multirow}
\usepackage{siunitx}
\usepackage[hidelinks]{hyperref}

\begin{document}

\title{Automatic Speech Recognition in the Modern Era: Architectures, Training, and Evaluation}

\author{
    \IEEEauthorblockN{Md Shamse Tabrej, Kabbojit Jit Deb, Md. Azizul Hakim, Shaonti Goswami }
    \IEEEauthorblockA{
        Department of Computer Science, Delhi Technological University (DTU), India\\
        Emails: \{mdshamsetabrej@gmail.com, kabbodeb@gmail.com, ahanik94@gmail.com, goswamishaonti@gmail.com\}
    }
    \and
    \IEEEauthorblockN{Md. Nayeem}
    \IEEEauthorblockA{
        National University of Bangladesh\\
        Email: md.nayeem6363@gmail.com
    }
}

\maketitle

\begin{abstract}
Automatic Speech Recognition (ASR) has undergone a profound transformation over the past decade, driven by advances in deep learning. This survey provides a comprehensive overview of the modern era of ASR, charting its evolution from traditional hybrid systems, such as Gaussian Mixture Model-Hidden Markov Models (GMM-HMMs) and Deep Neural Network-HMMs (DNN-HMMs), to the now-dominant end-to-end neural architectures. We systematically review the foundational end-to-end paradigms: Connectionist Temporal Classification (CTC), attention-based encoder-decoder models, and the Recurrent Neural Network Transducer (RNN-T), which established the groundwork for fully integrated speech-to-text systems. We then detail the subsequent architectural shift towards Transformer and Conformer models, which leverage self-attention to capture long-range dependencies with high computational efficiency. A central theme of this survey is the parallel revolution in training paradigms. We examine the progression from fully supervised learning, augmented by techniques like SpecAugment, to the rise of self-supervised learning (SSL) with foundation models such as wav2vec 2.0, which drastically reduce the reliance on transcribed data. Furthermore, we analyze the impact of large-scale, weakly supervised models like Whisper, which achieve unprecedented robustness through massive data diversity. The paper also covers essential ecosystem components, including key datasets and benchmarks (e.g., LibriSpeech, Switchboard, CHiME), standard evaluation metrics (e.g., Word Error Rate), and critical considerations for real-world deployment, such as streaming inference, on-device efficiency, and the ethical imperatives of fairness and robustness. We conclude by outlining open challenges and future research directions.
\end{abstract}

\begin{IEEEkeywords}
Automatic speech recognition; end-to-end; self-supervised learning; streaming; robustness; evaluation
\end{IEEEkeywords}

\section{Introduction}
Automatic Speech Recognition (ASR), the task of converting spoken language into text, is a cornerstone of modern human-computer interaction, powering applications from voice assistants and dictation software to in-car control systems and automated transcription services \cite{park2019specaugment}. The field has witnessed remarkable progress, evolving from early systems rooted in statistical methods to the highly performant deep learning models of today \cite{hinton2012deep, graves2013speech}.

Historically, ASR systems were constructed from a complex pipeline of independently trained components: an acoustic model (AM), typically a Gaussian Mixture Model-Hidden Markov Model (GMM-HMM), a pronunciation model (lexicon), and a language model (LM) \cite{povey2011kaldi}. The advent of deep neural networks (DNNs) replaced GMMs in hybrid DNN-HMM systems, yielding significant accuracy improvements \cite{hinton2012deep}. However, this modular design was complex, required domain expertise for each component, and suffered from error propagation between stages.

The modern era of ASR is defined by a paradigm shift towards end-to-end (E2E) models \cite{graves2006connectionist, chan2016listen}. These systems learn a direct mapping from acoustic features to text using a single, jointly optimized neural network, simplifying the training process and often improving performance. This survey focuses on the architectures, training paradigms, and evaluation methodologies that characterize this modern, E2E-centric landscape.

A second, equally transformative trend has been the shift in data paradigms. While supervised learning on large, transcribed corpora drove early deep learning successes, the field has increasingly moved towards methods that leverage vast quantities of unlabeled audio data. Self-supervised learning (SSL) frameworks, most notably wav2vec 2.0 \cite{baevski2020wav2vec}, have demonstrated the ability to learn powerful speech representations from raw audio, which can then be fine-tuned with minimal labeled data to achieve state-of-the-art results. Concurrently, large-scale weak supervision, exemplified by OpenAI's Whisper model \cite{radford2023robust}, has shown that training on massive, diverse, and multilingual web-scale data can produce a single model with remarkable zero-shot robustness across numerous domains and conditions.

This survey aims to provide a structured and comprehensive overview of these developments. The main contributions are as follows:
\begin{itemize}
    \item A systematic review of the dominant E2E ASR architectures, including Connectionist Temporal Classification (CTC), attention-based encoder-decoders (AED), Recurrent Neural Network Transducers (RNN-T), and the current state-of-the-art Transformer and Conformer models.
    \item A detailed analysis of modern training paradigms, tracing the evolution from supervised methods and data augmentation to the transformative impact of self-supervised and weakly-supervised learning.
    \item A comprehensive summary of the key datasets, benchmarks, and evaluation metrics that are used to measure and compare ASR system performance.
    \item An examination of practical deployment challenges, such as streaming inference for real-time applications and model efficiency for on-device processing, and a discussion of critical ethical considerations, including system robustness, fairness, and data privacy.
\end{itemize}

By synthesizing these interconnected topics, this paper serves as a reference for researchers and practitioners seeking to understand the current state and future trajectory of automatic speech recognition.

\section{Related Work}
Numerous surveys have chronicled the progress of ASR over the years. Early reviews focused on the statistical foundations of GMM-HMM systems and the initial integration of neural networks \cite{yu2016automatic}. More recent surveys have documented the transition to deep learning, often comparing hybrid DNN-HMM systems with the first generation of E2E models \cite{watanabe2018espnet}. For instance, Saon et al. provided an overview of English conversational speech recognition, highlighting the challenges of the Switchboard task \cite{saon2017english}. A comprehensive survey by Battenberg et al. explored various E2E approaches, comparing the merits of CTC and attention-based models \cite{battenberg2017exploring}.

However, the field has advanced at a rapid pace since these publications. The consolidation of the Transformer architecture \cite{vaswani2017attention} and its speech-specific variant, the Conformer \cite{gulati2020conformer}, has established a new baseline for acoustic modeling. Simultaneously, the rise of self-supervised learning \cite{baevski2020wav2vec} and large-scale pre-training \cite{radford2023robust} has fundamentally altered the data requirements and training methodologies for state-of-the-art ASR. These developments have rendered many earlier surveys incomplete.

This paper distinguishes itself by focusing specifically on this contemporary landscape, which we define as the period marked by the dominance of Transformer-based architectures and the widespread adoption of self-supervised and weakly-supervised training. We position our work to complement, rather than replace, prior surveys by providing a holistic narrative that integrates four key themes that are often treated in isolation:
\begin{enumerate}
    \item \textbf{Architectures:} We focus on the E2E models that define current research and production systems, from foundational concepts to the Conformer.
    \item \textbf{Training Paradigms:} We place significant emphasis on SSL and large-scale weak supervision, which represent the most significant shift in ASR development in recent years.
    \item \textbf{Deployment and Evaluation:} We connect architectural and training choices to real-world constraints like latency and computational cost, providing a practical perspective.
    \item \textbf{Ethics and Robustness:} We address the growing importance of evaluating ASR systems not just for accuracy, but also for their robustness in noisy conditions and their fairness across diverse speaker populations.
\end{enumerate}
By weaving these threads together, this survey provides a cohesive and up-to-date reference for the modern era of ASR.

\section{Architectures and Decoding}
\begin{figure}[h]
\centering
 \includegraphics[width=\linewidth]{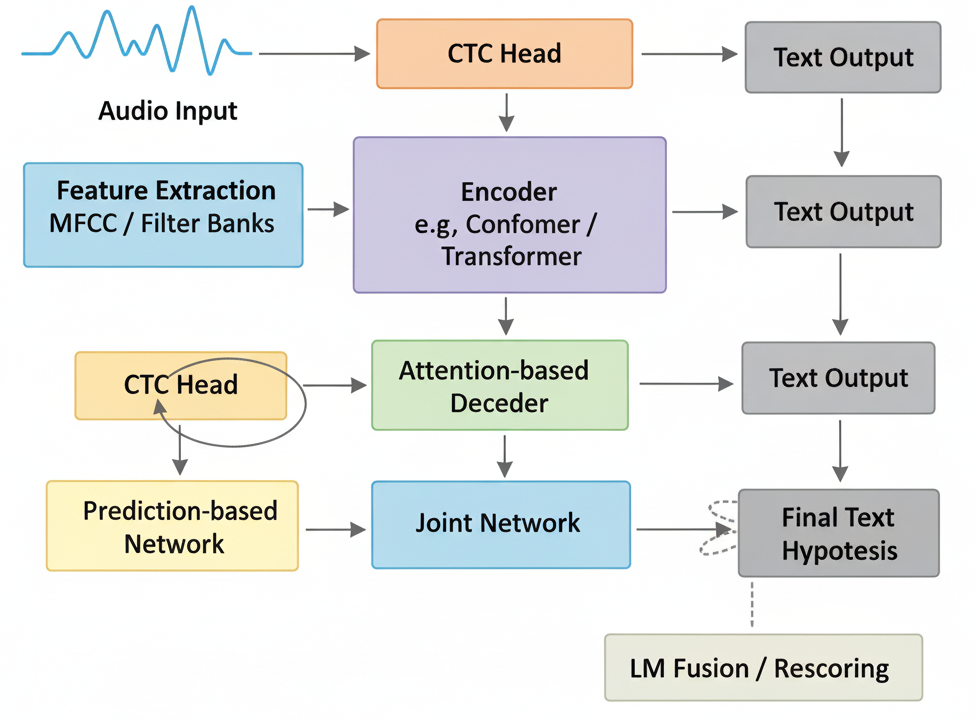}
\caption{A generalized overview of modern end-to-end ASR pipelines. An acoustic encoder processes audio features, and one of several decoding mechanisms (CTC, attention-based decoder, or RNN-T joint network) generates the output text. An external language model can optionally be used to rescore hypotheses.}
\label{fig:pipeline}
\end{figure}

The transition from hybrid to E2E systems was driven by the development of neural network architectures capable of learning the alignment between a variable-length audio input sequence and a variable-length text output sequence. This section details the three foundational E2E paradigms---CTC, AED, and RNN-T---and their evolution into the current state-of-the-art Transformer and Conformer models. A canonical overview of the modern ASR pipeline is shown in Fig. \ref{fig:pipeline}.

\subsection{Connectionist Temporal Classification (CTC)}
Introduced by Graves et al. \cite{graves2006connectionist}, Connectionist Temporal Classification (CTC) was a pioneering approach to E2E ASR. It addresses the alignment problem by augmenting the output vocabulary (e.g., characters) with a special `blank` token, denoted as $\epsilon$. The acoustic model, typically a Recurrent Neural Network (RNN), processes the input audio sequence $X = (x_1, \dots, x_T)$ and produces a probability distribution over this augmented vocabulary for each time step, resulting in an output matrix of size $T \times (|\mathcal{V}|+1)$, where $|\mathcal{V}|$ is the size of the original vocabulary.

The core idea of CTC is to treat any output sequence that reduces to the correct target label sequence after removing repeated characters and `blank` tokens as a valid alignment. For example, if the target is "CAT", paths like `C-A-A-T`, `CC-A-TT`, and `-C-A-T-` (where `-` is the blank token) are all considered correct. The CTC loss function is the negative log probability of the true label sequence, computed by efficiently summing the probabilities of all valid alignment paths using a dynamic programming algorithm similar to the forward-backward algorithm used in HMMs \cite{hannun2017sequence}.

The primary strength of CTC is its non-autoregressive nature. The prediction at each time step is conditionally independent of other predictions given the acoustic input. This allows the probabilities for all time steps to be computed in parallel, making it fast for both training and inference. However, this same independence assumption is its main weakness; CTC models do not inherently learn linguistic constraints or a language model, as the probability of a character does not depend on the previously emitted characters. Consequently, CTC-based systems often require a strong external language model during decoding to achieve competitive performance \cite{graves2014towards}.

\subsection{Attention-Based Encoder-Decoder (AED)}
Attention-based encoder-decoder (AED) models, also known as attention-based sequence-to-sequence models, offer an alternative solution to the alignment problem. Popularized in ASR by "Listen, Attend, and Spell" (LAS) \cite{chan2016listen}, these models consist of two main components: an encoder and a decoder.

\begin{enumerate}
    \item \textbf{Encoder (Listener):} This component functions similarly to the network in a CTC-based model. It takes the sequence of acoustic features and maps it to a higher-level sequence of hidden representations, $H = (h_1, \dots, h_{T'})$. The encoder is often a pyramidal RNN (e.g., BiLSTM), which progressively reduces the temporal resolution to create a more compact representation \cite{chan2016listen}.
    \item \textbf{Decoder (Speller):} This is an autoregressive model that generates the output sequence $Y = (y_1, \dots, y_U)$ one token at a time. At each step $i$, the decoder computes a context vector $c_i$ via an attention mechanism over the encoder outputs $H$. This context vector represents the most relevant acoustic information for predicting the current token $y_i$. The probability of $y_i$ is then conditioned on the previous tokens and the context vector: $P(y_i | y_{1..i-1}, X)$.
\end{enumerate}

The attention mechanism is the key component, as it explicitly learns a soft alignment between the input audio frames and the output text tokens. Unlike CTC, AED models are not conditionally independent; they directly model the probability of the entire output sequence using the chain rule, thereby learning an implicit language model from the training data. This often leads to better performance than CTC models without an external LM. The main drawback of AEDs is their autoregressive nature, which makes decoding inherently sequential and slower than CTC.

\subsection{Recurrent Neural Network Transducer (RNN-T)}
The Recurrent Neural Network Transducer (RNN-T), first proposed by Graves \cite{graves2012sequence}, combines the strengths of both CTC and AED models. It has become a dominant architecture for production-grade streaming ASR systems \cite{he2019streaming}. The RNN-T architecture consists of three components:

\begin{enumerate}
    \item \textbf{Acoustic Encoder:} Similar to CTC and AED models, it processes the acoustic input $X$ to produce a sequence of representations $f_t$.
    \item \textbf{Prediction Network:} An autoregressive RNN that processes the previously predicted non-blank labels $y_{1..u-1}$ to produce a representation $g_u$. This component acts as an internal language model.
    \item \textbf{Joint Network:} A feed-forward network that combines the outputs of the acoustic encoder ($f_t$) and the prediction network ($g_u$) to produce a probability distribution over the augmented vocabulary (labels + `blank`).
\end{enumerate}

At each time step $(t, u)$, the joint network computes $P(y | t, u)$. If a non-blank label is emitted, the label index $u$ is incremented. If a `blank` is emitted, the acoustic frame index $t$ is incremented. This process continues until the end of the audio sequence is reached. Like CTC, the RNN-T loss function is calculated by summing over all possible alignments using a forward-backward algorithm.

The RNN-T architecture is naturally suited for streaming because the computation at each step depends only on the current acoustic frame and the previously emitted labels. It does not need to see the entire audio sequence, making it ideal for low-latency applications. It also incorporates an internal language model via the prediction network, overcoming a key limitation of CTC.

\subsection{Transformer and Conformer Architectures}
While RNNs were the backbone of early E2E models, they have been largely superseded by Transformer-based architectures \cite{vaswani2017attention}, which use self-attention mechanisms to model sequential dependencies.

The \textbf{Speech-Transformer} \cite{dong2018speech} was one of the first works to successfully adapt the Transformer for ASR. The encoder and decoder are stacks of multi-head self-attention and feed-forward layers. Self-attention allows the model to weigh the importance of all other frames in the sequence when encoding a specific frame, enabling it to capture long-range dependencies more effectively than RNNs. The removal of recurrence also allows for significantly more parallelization during training.

The \textbf{Conformer} architecture \cite{gulati2020conformer} further improved upon the Transformer by explicitly combining self-attention with convolutions to model both global and local context, respectively. A Conformer block typically consists of a feed-forward module, a multi-head self-attention module, a convolution module, and a final feed-forward module, arranged in a "macaron-like" structure. This hybrid approach has proven extremely effective, and Conformers, often paired with a CTC or RNN-T loss function, represent the state of the art for many ASR benchmarks \cite{gulati2020conformer}.

\subsection{Decoding and Language Model Fusion}
Inference in E2E models, especially autoregressive ones, involves a search algorithm to find the most probable output sequence. Simple greedy decoding (picking the most likely token at each step) is fast but suboptimal. \textbf{Beam search} is more commonly used, where a fixed number of candidate hypotheses (the "beam") are maintained at each step.

To further improve accuracy, an external language model (LM), trained on large text corpora, can be integrated during decoding. This is particularly crucial for CTC-based models. A common technique is \textbf{shallow fusion}, where the LM score is linearly interpolated with the acoustic model score at each step of the beam search:
$$\log P(y) = \log P_{AM}(y|X) + \alpha \log P_{LM}(y)$$
where $\alpha$ is a tunable weight. More advanced techniques like cold fusion integrate the LM directly into the model's training process.

\subsection{Open-Source Toolkits}
The rapid progress in ASR has been facilitated by powerful open-source toolkits. \textbf{Kaldi} \cite{povey2011kaldi} has been the long-standing standard for building traditional hybrid ASR systems. For modern E2E research, \textbf{ESPnet} \cite{watanabe2018espnet} provides a comprehensive and unified framework for a wide range of architectures and tasks. \textbf{Fairseq} \cite{ott2019fairseq}, developed by Facebook AI, is a general sequence modeling toolkit that provides reference implementations for influential models like wav2vec 2.0. These toolkits provide reproducible recipes for major benchmarks, lowering the barrier to entry and fostering innovation.

\section{Training Paradigms}
The performance of modern ASR architectures is inextricably linked to the paradigms used for their training. While fully supervised learning remains a foundation, recent years have seen a decisive shift towards techniques that either augment existing labeled data or leverage vast quantities of unlabeled data to build more robust and accurate models.

\subsection{Supervised Learning and Data Augmentation}
The conventional approach to training ASR models involves supervised learning on a dataset of paired audio utterances and their corresponding ground-truth transcriptions. The model is trained to minimize a loss function (e.g., CTC, cross-entropy) that measures the discrepancy between its predictions and the reference text.

To improve model robustness and generalization, data augmentation is a critical step. While traditional methods involve adding noise or reverberation to raw audio, a highly effective and computationally inexpensive technique is \textbf{SpecAugment} \cite{park2019specaugment}. Applied directly to the log-mel spectrogram features, SpecAugment consists of three transformations:
\begin{itemize}
    \item \textbf{Time Warping:} A random deformation of the spectrogram along the time axis.
    \item \textbf{Frequency Masking:} Masking out a random contiguous block of frequency channels. This makes the model more robust to partial loss of frequency information.
    \item \textbf{Time Masking:} Masking out a random contiguous block of time steps. This encourages the model to learn from the surrounding context and improves robustness to occlusions or short bursts of noise.
\end{itemize}
SpecAugment has become a standard component in ASR training pipelines, as it significantly reduces WER without requiring any additional data or complex feature engineering \cite{park2019specaugment}.

\subsection{Regularization and Curriculum Learning}
\textbf{Label smoothing} is a regularization technique that prevents a model from becoming overconfident in its predictions \cite{szegedy2016rethinking}. Instead of using one-hot encoded target labels (where the correct class has a probability of 1.0 and all others have 0.0), the target distribution is "smoothed" by distributing a small probability mass $\alpha$ across all classes. This encourages the model to produce softer output distributions, which has been shown to improve both accuracy and model calibration \cite{muller2019does}.

\textbf{Curriculum learning} is a training strategy inspired by human learning, where the model is first presented with easier examples and the difficulty is gradually increased \cite{bengio2009curriculum}. In ASR, this can be implemented by first training the model on high-quality, clean speech and progressively introducing samples with lower signal-to-noise ratios (SNRs) \cite{braun2017curriculum}. This staged approach can help the model converge faster and achieve better noise robustness.

\begin{table}[t]
\centering
\caption{Common Training Practices in Modern ASR.}
\label{tab:train}
\begin{tabular}{l p{2.7cm} p{2.7cm}}
\toprule
\textbf{Practice} & \textbf{Purpose} & \textbf{Notes} \\
\midrule
SpecAugment & Data augmentation for improved robustness & Applies time warping, frequency masking, and time masking to spectrograms \cite{park2019specaugment}. \\
Label Smoothing & Regularization to prevent overconfidence & Replaces hard one-hot labels with a smoothed distribution. Typical $\alpha \in [0.05, 0.2]$ \cite{szegedy2016rethinking}. \\
LM Fusion & Accuracy improvement via external knowledge & Linearly combines AM and LM scores during beam search decoding (shallow fusion) \cite{kannan2018analysis}. \\
Self-Supervised Pre-training & Learn representations from unlabeled data & e.g., wav2vec 2.0 uses a contrastive task on masked latent audio \cite{baevski2020wav2vec}. \\
Curriculum Learning & Improved convergence and robustness & Training progresses from easy examples (high SNR) to hard examples (low SNR) \cite{braun2017curriculum}. \\
\bottomrule
\end{tabular}
\end{table}

\subsection{Self-Supervised Learning (SSL)}
The most significant shift in ASR training has been the rise of self-supervised learning (SSL), which enables models to learn powerful representations from unlabeled audio data, dramatically reducing the need for expensive transcribed corpora. The seminal work in this area is \textbf{wav2vec 2.0} \cite{baevski2020wav2vec}.

The wav2vec 2.0 framework involves a two-stage process:
\begin{enumerate}
    \item \textbf{Pre-training:} A large model, typically a Transformer, is trained on thousands of hours of unlabeled speech. The raw audio waveform is first passed through a convolutional feature encoder to obtain a sequence of latent representations. A random subset of these representations is then masked. The model is trained on a contrastive task: for each masked time step, it must identify the correct \textit{quantized} version of its latent representation from a set of distractors. This forces the model to learn high-level contextualized representations of the speech signal.
    \item \textbf{Fine-tuning:} After pre-training, a small, randomly initialized linear layer is added on top of the Transformer encoder, and the entire model is fine-tuned on a small amount of labeled data for a specific ASR task (e.g., transcribing English).
\end{enumerate}
The power of this approach is its data efficiency. Baevski et al. demonstrated that a wav2vec 2.0 model pre-trained on 53k hours of unlabeled audio could achieve state-of-the-art results on the LibriSpeech benchmark using as little as ten minutes of labeled data for fine-tuning \cite{baevski2020wav2vec}. This has democratized the development of high-performance ASR for low-resource languages and domains where transcribed data is scarce.

\subsection{Large-Scale Weak Supervision}
An alternative approach to leveraging large datasets is weak supervision. Instead of a carefully designed self-supervised objective, this paradigm relies on the sheer scale and diversity of existing, albeit noisy, audio-transcript pairs available on the web. The most prominent example is OpenAI's \textbf{Whisper} model \cite{radford2023robust}.

Whisper is a large Transformer-based encoder-decoder model trained on an enormous dataset of 680,000 hours of multilingual and multitask audio collected from the internet. The "weak supervision" comes from the fact that these transcripts are not guaranteed to be perfectly accurate. However, by training on such a vast and diverse dataset, the model learns to be inherently robust to a wide range of acoustic conditions, including background noise, different accents, and various speaking styles. A key finding of this work is that a single, large model trained this way can achieve remarkable \textit{zero-shot} performance on many standard ASR benchmarks, often competitive with systems that were specifically fine-tuned on those datasets \cite{radford2023robust}. This approach suggests that data scale and diversity can be a powerful substitute for meticulously curated datasets and complex training objectives.

\section{Datasets and Benchmarks}
The development and evaluation of ASR systems are grounded in standardized datasets that serve as common benchmarks for the research community. These datasets vary widely in size, domain, recording conditions, and linguistic content, allowing for the assessment of different aspects of system performance.

\begin{figure}[h]
\centering
\includegraphics[width=\linewidth]{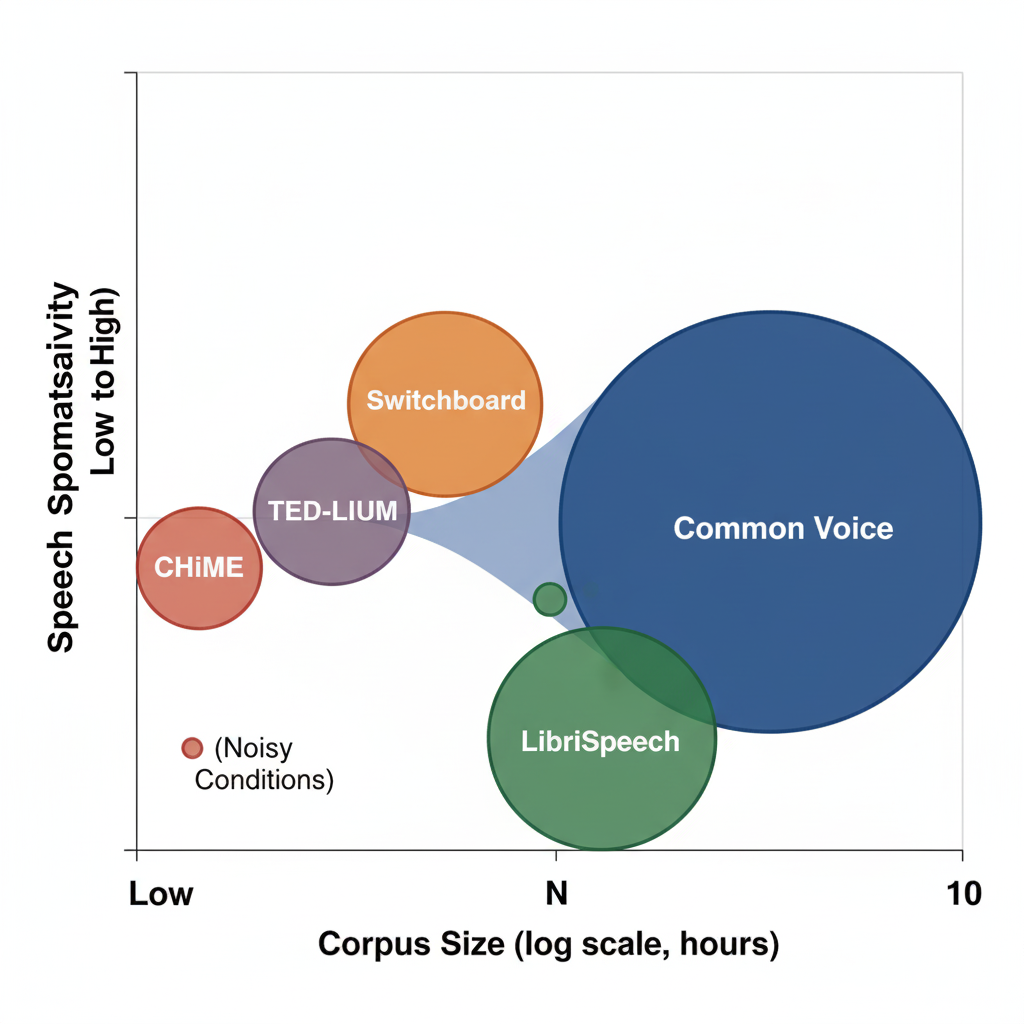}
\caption{A conceptual landscape of prominent ASR datasets, plotted by corpus size versus the degree of speech spontaneity. Bubble size indicates multilingual coverage.}
\label{fig:data}
\end{figure}

\subsection{LibriSpeech}
The LibriSpeech corpus \cite{panayotov2015librispeech} is arguably the most widely used benchmark for ASR in the academic community. It consists of approximately 1000 hours of 16 kHz read English speech derived from audiobooks from the LibriVox project. The data is segmented, aligned, and organized into standardized training, development, and test sets. A key feature is the division of its evaluation sets into `test-clean` and `test-other` subsets, which represent recordings from speakers who are easier and harder to recognize, respectively. This allows for a nuanced evaluation of model performance under both ideal and more challenging (but still clean) acoustic conditions. Its permissive CC BY 4.0 license has contributed to its widespread adoption.

\subsection{Switchboard}
The Switchboard corpus \cite{godfrey1992switchboard} is a foundational dataset for conversational telephone speech. It contains approximately 300 hours of spontaneous, two-sided telephone conversations between 543 speakers of American English. Participants were given a topic to discuss, resulting in natural, unscripted speech complete with disfluencies, hesitations, and interruptions. The audio is sampled at 8 kHz, reflecting its telephone channel origin. Despite its age, Switchboard remains a challenging and relevant benchmark for evaluating ASR performance on spontaneous conversational speech. It is distributed by the Linguistic Data Consortium (LDC) and requires a license for use.

\subsection{TED-LIUM}
The TED-LIUM corpus is derived from TED Talks, offering a large collection of speech from a diverse set of speakers on a wide range of topics. The third release, TED-LIUM 3 \cite{hernandez2018tedlium3}, contains 452 hours of 16 kHz English speech. The domain is prepared, semi-spontaneous speech from public talks, which presents different challenges from read audiobooks or telephone conversations, including diverse accents, variable speaking rates, and occasional audience noise. It is a valuable resource for training and evaluating systems intended for lecture or presentation transcription.

\subsection{CHiME Challenges}
The CHiME (Computational Hearing in Multisource Environments) challenge series is designed to spur research in robust ASR in noisy, real-world environments \cite{barker2018fifth}. The datasets feature distant-microphone recordings in challenging acoustic settings. For example, the CHiME-6 challenge \cite{watanabe2020chime} uses data from dinner parties recorded in real homes using multiple microphone arrays. This captures complex acoustic scenes with high levels of background noise, reverberation, and overlapping speech from multiple speakers. The CHiME datasets are essential for evaluating the true robustness of ASR systems beyond clean, laboratory-like conditions.

\subsection{Common Voice}
The Common Voice project by Mozilla is a massive, crowdsourced initiative to create a free and publicly available speech dataset for a wide range of languages \cite{ardila2020common}. Contributors record their voices by reading sentences from a text corpus, and other users validate the recordings. The dataset's scale is its primary advantage; recent releases contain thousands of validated hours across over 100 languages \cite{commonvoice17}. Furthermore, it includes optional demographic metadata from contributors, such as age, gender, and accent. This makes Common Voice an invaluable resource not only for building multilingual ASR systems but also for studying and mitigating demographic bias and fairness issues in speech technology. The data is released under the permissive CC0 license.

\begin{table}[t]
\centering
\caption{Characteristics of Representative ASR Datasets.}
\label{tab:data}
\sisetup{table-format=4.0, table-number-alignment=center}
\begin{tabular}{l S[table-format=5.0] S l}
\toprule
\textbf{Dataset} & {\textbf{Hours}} & {\textbf{Speakers}} & \textbf{Domain / Notes} \\
\midrule
LibriSpeech \cite{panayotov2015librispeech} & 960 & 2484 & Read English audiobooks \\
Switchboard \cite{godfrey1992switchboard} & 300 & 543 & English telephone conversations \\
TED-LIUM 3 \cite{hernandez2018tedlium3} & 452 & 2351 & English talks, diverse accents \\
CHiME-6 \cite{watanabe2020chime} & 50 & 20 & Noisy, distant-mic conversations \\
Common Voice 17.0 \cite{commonvoice17} & {>20000} & {>100k} & Crowdsourced, 124 languages \\
\bottomrule
\end{tabular}
\end{table}

\section{Evaluation and Reported Results}
Evaluating the performance of ASR systems requires well-defined metrics and standardized benchmarks. This section outlines the primary metrics used in the field and presents a snapshot of reported results for leading models on the LibriSpeech benchmark.

\subsection{Evaluation Metrics}
The most common metric for ASR accuracy is the \textbf{Word Error Rate (WER)}, which measures the dissimilarity between the ASR system's hypothesis and a ground-truth reference transcript. It is calculated based on the Levenshtein distance at the word level:
$$\text{WER} = \frac{S + D + I}{N}$$
where $S$ is the number of substitutions, $D$ is the number of deletions, $I$ is the number of insertions, and $N$ is the total number of words in the reference transcript. A lower WER indicates higher accuracy. For languages that are not whitespace-delimited or are highly agglutinative, the \textbf{Character Error Rate (CER)} is often used instead, calculated analogously at the character level.

While WER and CER measure accuracy, they do not capture the real-time performance of a system. For streaming applications, two key metrics are:
\begin{itemize}
    \item \textbf{Latency:} The time delay (in milliseconds) from when a word is spoken to when its transcript is finalized.
    \item \textbf{Real-Time Factor (RTF):} The ratio of the processing time to the duration of the audio. An RTF less than 1.0 is necessary for a system to keep up with a live audio stream.
\end{itemize}
A comprehensive evaluation of an ASR system requires assessing the trade-off between accuracy (WER) and these performance metrics. Statistical significance tests, such as matched pairs tests, are often used to confirm that differences in WER between systems are not due to chance.

\subsection{Reported Results on LibriSpeech}
The LibriSpeech benchmark serves as a standard for comparing the performance of different ASR architectures and training methods. Table \ref{tab:sota} provides an illustrative snapshot of published WERs for several prominent models on the `test-clean` and `test-other` evaluation sets. These results highlight the significant progress made by modern E2E systems. Conformer-based models and SSL-based models like wav2vec 2.0, when fully supervised or fine-tuned on the 960-hour training set, have pushed performance to very low error rates. The Whisper model is notable for achieving competitive performance in a zero-shot setting, without any fine-tuning on LibriSpeech, demonstrating the power of large-scale, diverse pre-training.

\begin{table*}[t]
\centering
\caption{Published Word Error Rates (\%) on the LibriSpeech Benchmark (Illustrative Snapshot).}
\label{tab:sota}
\begin{tabular}{l c c c l}
\toprule
\textbf{Model} & \textbf{test-clean} & \textbf{test-other} & \textbf{Latency/RTF} & \textbf{Source} \\
\midrule
Conformer-T (non-streaming, with LM) & 1.9 & 3.9 & --- & Gulati et al. (2020) \cite{gulati2020conformer} \\
wav2vec 2.0 (LARGE, fine-tuned 960h, with LM) & 1.8 & 3.3 & --- & Baevski et al. (2020) \cite{baevski2020wav2vec} \\
Whisper (large-v2, zero-shot) & 2.7 & 5.0 & --- & Radford et al. (2023) \cite{radford2023robust} \\
Streaming Conformer (960ms chunk) & 2.72 & 6.47 & Streaming & SpeechBrain \cite{speechbrain-streaming-conformer} \\
\bottomrule
\end{tabular}
\end{table*}

\begin{figure}[t]
\centering
\includegraphics[width=\linewidth]{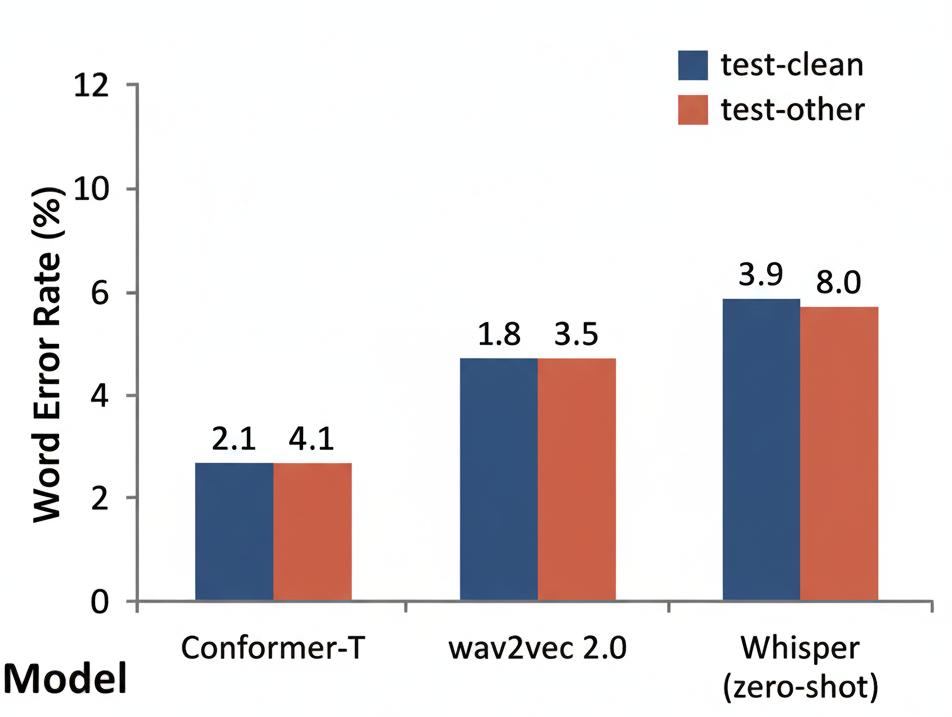}
\caption{Comparison of published Word Error Rates (WERs) for leading ASR models on the LibriSpeech `test-clean` and `test-other` sets. Lower is better.}
\label{fig:wer}
\end{figure}

\section{Streaming, On-Device, and Efficiency}
While achieving high accuracy on offline benchmarks is a primary research goal, deploying ASR in real-world applications imposes strict constraints on latency, computational resources, and privacy. This section explores the techniques used to build streaming, on-device, and efficient ASR systems.

\subsection{Streaming ASR}
Many ASR applications, such as live captioning and voice assistants, require the system to transcribe speech in real time as it is being spoken. This necessitates a streaming inference capability, where the model processes audio in small chunks and produces output with minimal delay.

Non-streaming (or full-context) models, which require the entire utterance before starting transcription, are unsuitable for this task. Architectures like the RNN-Transducer are naturally streamable due to their monotonic alignment property \cite{graves2012sequence}. For Transformer and Conformer models, which rely on self-attention over the entire sequence, streaming is enabled through techniques like \textbf{chunked attention} \cite{miao2020transformer}. Here, the input audio is divided into fixed-size chunks. When processing a given chunk, the self-attention mechanism is restricted to attend only to the current chunk and a limited number of preceding chunks (the "left context"). This prevents the model from accessing future audio frames, enabling causal, chunk-by-chunk processing.

An effective streaming system also requires \textbf{voice activity detection (VAD)} and \textbf{endpointing}. VAD identifies segments of audio that contain speech, while endpointing determines when a speaker has likely finished their utterance, allowing the system to finalize the current hypothesis and reset for the next one.

\begin{figure}[h]
\centering
\includegraphics[width=\linewidth]{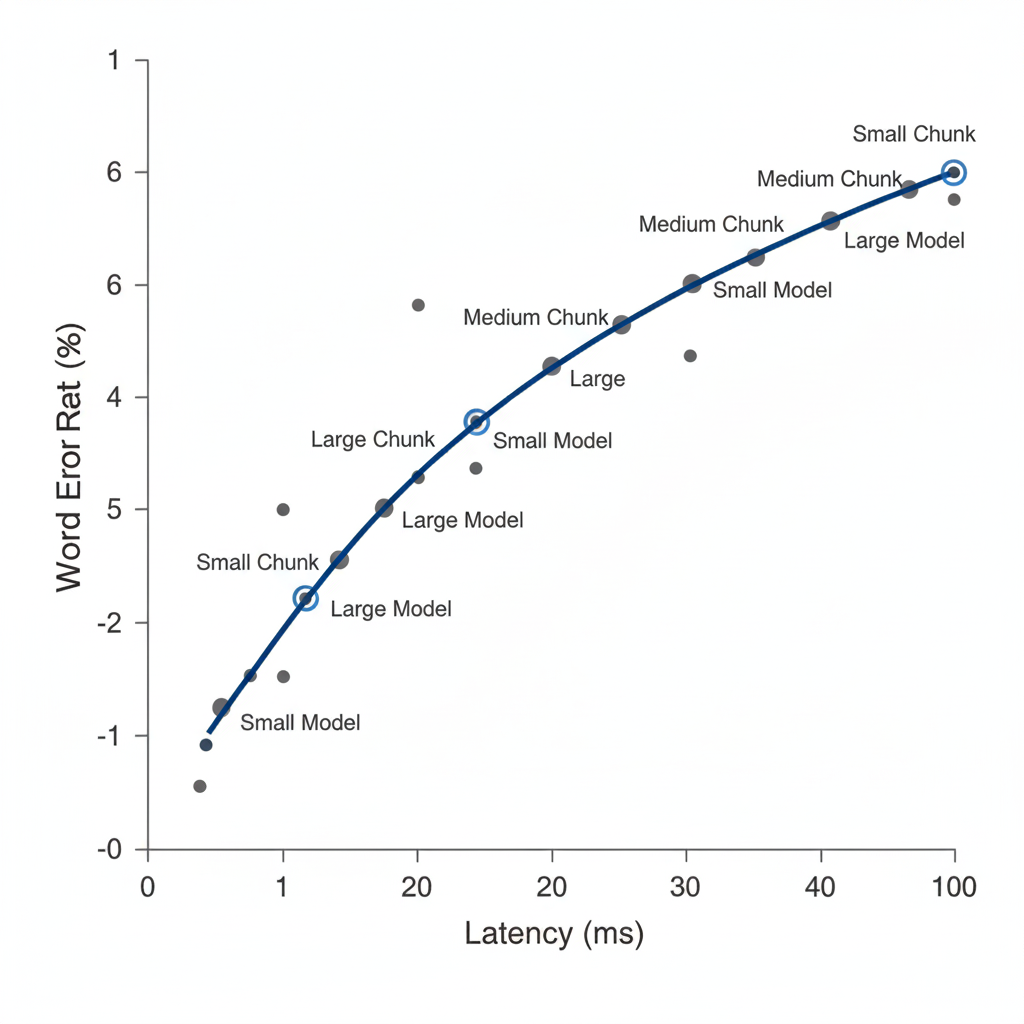}
\caption{An illustration of the fundamental trade-off between latency and accuracy in streaming ASR systems. The Pareto frontier represents the set of optimal configurations where one metric cannot be improved without degrading the other.}
\label{fig:latency}
\end{figure}

\subsection{On-Device Processing and Efficiency}
Running ASR models directly on an end-user's device (e.g., a smartphone or smart speaker) offers significant advantages over cloud-based processing:
\begin{itemize}
    \item \textbf{Privacy:} Voice data is not sent to a third-party server, mitigating privacy concerns.
    \item \textbf{Latency:} Eliminates network round-trip time, resulting in a more responsive user experience.
    \item \textbf{Availability:} The system can function without an internet connection.
\end{itemize}
However, on-device deployment is challenging due to the limited computational power and memory of edge devices. State-of-the-art ASR models can have hundreds of millions of parameters and require significant computational resources. To make these models feasible for on-device use, \textbf{model compression} techniques are essential. The two most common methods are:

\begin{enumerate}
    \item \textbf{Quantization:} This involves reducing the numerical precision of the model's weights and activations. Instead of using 32-bit floating-point numbers, weights can be represented using 16-bit floats or, more commonly, 8-bit integers (INT8). This reduces the model's memory footprint and can significantly accelerate inference on hardware with specialized support for lower-precision arithmetic \cite{han2016deep}.
    \item \textbf{Pruning:} This technique involves removing redundant or less important connections (weights) from the neural network. In \textit{unstructured pruning}, individual weights with small magnitudes are set to zero, creating sparse weight matrices. In \textit{structured pruning}, entire groups of weights, such as filters or network layers, are removed. While unstructured pruning can achieve higher compression rates, structured pruning is often more effective at achieving practical speedups on modern hardware designed for dense matrix operations \cite{han2015learning}.
\end{enumerate}
These techniques, often used in combination, allow for the creation of compact and efficient models that can deliver high-quality ASR directly on edge devices, balancing the trade-off between accuracy, latency, and resource consumption as illustrated in Fig. \ref{fig:latency}.

\section{Robustness, Fairness, and Ethics}
As ASR technology becomes more pervasive, its societal impact grows, making it crucial to address issues of robustness, fairness, and ethics. A system that is highly accurate on benchmark data but fails in real-world noise or performs inequitably across different populations is not truly effective.

\subsection{Robustness to Acoustic Conditions}
Real-world audio is often corrupted by background noise (e.g., street sounds, music) and reverberation (reflections of sound in an enclosed space). These acoustic distortions can cause a significant degradation in ASR performance. The \textbf{CHiME challenges} \cite{barker2018fifth} have been instrumental in driving research on this problem. The datasets, recorded in challenging, noisy environments like cafes and homes, provide a benchmark for evaluating noise-robust ASR. Successful approaches often combine multi-microphone signal processing techniques (e.g., beamforming to isolate a target speaker) with multi-condition training, where the ASR model is trained on data that has been artificially corrupted with a wide variety of noise types and SNRs. The success of models like Whisper also demonstrates that training on massive and diverse real-world data is a powerful strategy for achieving inherent robustness \cite{radford2023robust}.

\subsection{Fairness and Demographic Bias}
A significant ethical challenge in ASR is performance bias. A growing body of research has demonstrated that ASR systems often exhibit substantially higher error rates for certain demographic groups compared to others \cite{koenecke2020racial}. These disparities can occur across various axes:
\begin{itemize}
    \item \textbf{Accent and Dialect:} Systems typically perform best on dominant, "standard" accents (e.g., Standard American English) and show significantly higher WERs for speakers with regional or non-native accents \cite{feng2024fairness}. One study found that commercial ASR systems had nearly double the error rate for speakers of African American Vernacular English (AAVE) compared to white speakers \cite{koenecke2020racial}.
    \item \textbf{Gender:} Some studies have reported higher WERs for female speakers than for male speakers, a gap often attributed to the underrepresentation of female speech in training data \cite{tatman2017gender}.
    \item \textbf{Age:} ASR systems tend to perform worse for children and older adults compared to middle-aged adults, whose speech patterns are typically better represented in training corpora \cite{feng2024fairness}.
\end{itemize}
The primary cause of these biases is the composition of the training data. If a particular demographic group is underrepresented, the model will not learn to recognize their speech patterns as effectively. This can lead to a cycle of exclusion, where poor performance discourages usage by marginalized groups, further limiting data collection and perpetuating the bias. Addressing this requires a concerted effort to collect more diverse and representative datasets, such as the Mozilla Common Voice corpus \cite{ardila2020common}, and to develop fairness-aware training and evaluation methodologies.

\subsection{Data Privacy and Licensing}
The use of ASR raises important privacy considerations. Cloud-based ASR services require users to send their voice data, which can be highly personal and sensitive, to remote servers for processing. This creates potential risks of data breaches or misuse. On-device ASR (as discussed in Section VII) is a powerful technical solution that mitigates these risks by keeping all data local. When data is collected for training, clear privacy policies and user consent are essential \cite{asr-privacy-policy}.

Furthermore, data licensing is a critical component of the ASR ecosystem. The progress of academic research depends heavily on open datasets with permissive licenses (e.g., CC0, CC BY) that allow for sharing and reproduction of results. Corpora like LibriSpeech and Common Voice have been pivotal in this regard, while commercially licensed datasets like Switchboard, though valuable, can present barriers to access.

\section{Open Challenges and Future Directions}
Despite immense progress, several significant challenges remain, pointing towards key areas for future research and development in ASR.

\subsection{Multilingual and Code-Switching ASR}
While ASR for high-resource languages like English has reached maturity, performance for the vast majority of the world's languages remains poor due to data scarcity. A major challenge is building robust \textbf{multilingual ASR} systems that can recognize speech from many languages using a single model. This often involves pooling data from multiple languages to leverage phonetic and linguistic similarities. Recent work on low-resource languages, such as Bangla, continues to explore effective architectures and training strategies in data-constrained settings \cite{bhadra2025deep}.

A more complex related problem is \textbf{code-switching}, where speakers alternate between two or more languages within a single conversation or even a single sentence. This is common in multilingual communities but poses a significant challenge for ASR systems, which must simultaneously manage multiple vocabularies, grammars, and phonetic inventories \cite{li2021multilingual}. Developing models that can seamlessly handle code-switching is a critical frontier for making ASR truly global.

\subsection{Personalization with Privacy}
Generic ASR models often struggle with user-specific vocabulary, such as contact names, local places, or technical jargon. \textbf{Personalization}, or adapting a model to an individual user's voice, accent, and vocabulary, can dramatically improve usability. However, this must be balanced with user privacy. Future research will likely focus on privacy-preserving adaptation techniques, such as on-device fine-tuning or federated learning, where models are updated on user data without the raw data ever leaving the device.

\subsection{Beyond WER Evaluation}
Word Error Rate, while a useful metric, is a blunt instrument. It treats all errors equally, yet the semantic impact of errors can vary dramatically; misrecognizing "two" as "to" is less severe than misrecognizing "accept" as "except". There is a growing need for evaluation metrics that go \textbf{beyond lexical accuracy} to measure semantic correctness, robustness against critical errors, and the overall usability of ASR output \cite{huang2023beyondwer}. Furthermore, developing label-free evaluation methods that can estimate ASR performance without ground-truth transcripts would enable assessment across a much wider range of real-world domains \cite{khurana2024label}.

\subsection{Related Speech Technologies}
The underlying architectures and training methods developed for ASR are also being applied to a broader set of speech processing tasks. For example, \textbf{Speech Emotion Recognition (SER)} aims to identify the emotional state of a speaker from their vocal cues. While it uses similar front-end processing, its goal is classification of affect rather than transcription of content. Research in SER explores different feature extraction methods, such as those based on wavelet transforms, and classification models like Support Vector Machines to identify emotions like happiness, sadness, or anger from speech signals \cite{hosain2024speech}. The synergy between ASR and other speech intelligence tasks represents a rich area for future cross-pollination of ideas.

\section{Conclusion}
The field of Automatic Speech Recognition has transitioned into a new era, characterized by the dominance of end-to-end neural architectures and a fundamental shift in how data is leveraged. The evolution from CTC and attention-based models to the powerful and efficient Conformer architecture has marked a period of rapid improvement in acoustic modeling. This architectural progress has been matched, and perhaps even surpassed, by a revolution in training paradigms. The move from purely supervised learning to self-supervised methods like wav2vec 2.0 and large-scale weak supervision with models like Whisper has unlocked the potential of massive unlabeled and web-scale datasets, drastically reducing the dependency on curated, transcribed corpora and imbuing models with unprecedented robustness.

This survey has provided a structured overview of this modern landscape. We have detailed the key E2E architectures, analyzed the training methodologies that power them, and summarized the datasets and metrics used to benchmark them. Furthermore, we have examined the critical, practical dimensions of the field, including the engineering challenges of building low-latency streaming and efficient on-device systems, and the pressing ethical need to ensure that ASR technology is robust, fair, and privacy-preserving.

While ASR has achieved near-human performance on several benchmarks, significant challenges remain. The frontiers of research are now pushing towards true multilingual understanding, seamless handling of code-switching, private and effective personalization, and more meaningful, beyond-WER evaluation. The continued convergence of advanced architectures, large-scale data, and a growing focus on real-world applicability promises to make speech technology even more capable and ubiquitous in the years to come.

\section*{Acknowledgment}
The authors would like to thank the open-source community and the creators of the public datasets cited in this work, whose contributions have been instrumental in advancing the field of automatic speech recognition.

\bibliographystyle{IEEEtran}
\bibliography{refs}

\appendices
\section{Reproducibility Notes}
The progress in modern ASR is heavily reliant on open and reproducible research. The following open-source toolkits are central to the development and evaluation of the models discussed in this survey:
\begin{itemize}
    \item \textbf{Kaldi:} \url{http://kaldi-asr.org/}
    \item \textbf{ESPnet:} \url{https://github.com/espnet/espnet}
    \item \textbf{Fairseq:} \url{https://github.com/facebookresearch/fairseq}
    \item \textbf{Hugging Face Transformers/Datasets:} \url{https://huggingface.co/}
    \item \textbf{SpeechBrain:} \url{https://speechbrain.github.io/}
\end{itemize}

\end{document}